\documentclass[10pt]{iopart}
\usepackage{iopams}  
\usepackage{graphicx}   
\usepackage{color}      
\usepackage{hyperref}   

\def\fetese{FeTe$_{1-x}$Se$_x$}
\def\feytese{Fe$_{1+y}$Te$_{1-x}$Se$_x$}

\newcommand{\bQ}{\mbox{\boldmath$Q$}}
\newcommand{\bq}{\mbox{\boldmath$q$}}

\begin{document}

\title[]{Magnetism and superconductivity in Fe$_{1+y}$Te$_{1-x}$Se$_x$}

\author{J. M. Tranquada,$^1$ Guangyong Xu,$^2$ and I. A. Zaliznyak$^1$}

\address{$^1$Condensed Matter \&\ Materials Science Division, Brookhaven National Laboratory, Upton, NY 11973-5000, USA}
\address{$^2$NIST Center for Neutron Research, National Institute of Standards and Technology, 100 Bureau Drive, Gaithersburg, Maryland 20899, USA}
\ead{jtran@bnl.gov}
\vspace{10pt}
\begin{indented}
\item[]\today
\end{indented}

\begin{abstract}
Neutron scattering has played a significant role in characterizing magnetic and structural correlations in Fe$_{1+y}$Te$_{1-x}$Se$_x$ and their connections with superconductivity.  Here we review several key aspects of the physics of iron chalcogenide superconductors where neutron studies played a key role. These topics include the phase diagram of Fe$_{1+y}$Te$_{1-x}$Se$_{x}$, where the doping-dependence of structural transitions can be understood from a mapping to the anisotropic random field Ising model. We then discuss orbital-selective Mott physics in the Fe chalcogenide series, where temperature-dependent magnetism in the parent material provided one of the earliest cases for orbital-selective correlation effects in a Hund's metal. Finally, we elaborate on the character of local magnetic correlations revealed by neutron scattering, its dependence on temperature and composition, and the connections to nematicity and superconductivity.
\end{abstract}

%
%
%
%
\ioptwocol

\section{Introduction}

The discovery of high-temperature superconductivity in iron-arsenide compounds came as a very pleasant surprise that has motivated a considerable amount of research \cite{hoso18}.  
Similar to copper-oxide superconductors, the superconductivity is typically induced by doping a parent compound to suppress antiferromagnetic order.  Different from cuprates, the parent compounds are generally metallic, which introduces the challenge of understanding the nature of an ``itinerant antiferromagnet''.  Given that sizable crystals can be grown for many of these compounds, this has been an excellent problem for experimental investigation by neutron scattering \cite{dai15}.

Along with many others \cite{wen11,wen15}, we have spent the last decade investigating the iron-chalcogenide system \feytese.  While the maximum superconducting transition temperature is only about 15~K, its simple crystal structure nevertheless yields surprisingly complex evolutions of the magnetic correlations with composition and temperature.  There is also new interest in the \fetese\ system following the recent reports of topological surface states \cite{zhan19,rame19} and Majorana bound states within vortices \cite{kong19}.

\begin{figure}[t]
\begin{center}
\includegraphics[width=0.95\columnwidth]{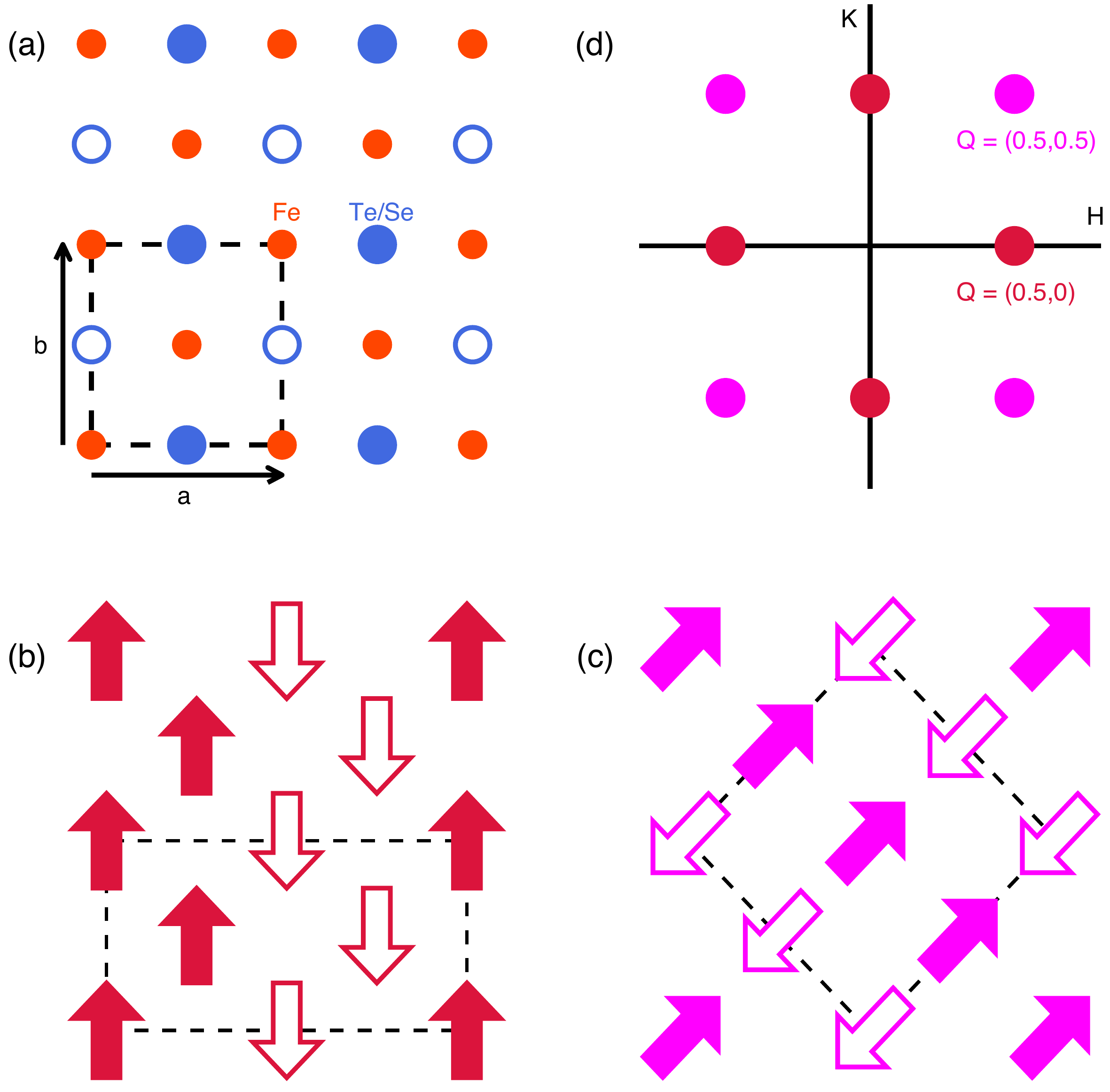}
\caption{\label{fg:struc} 
(a) Structure of one layer of \fetese: Fe atoms (red) form a square lattice: Te/Se atoms (blue) sit above (solid) or below (open) the Fe layer; $a$ and $b$ are the lattice parameters.  (b) Spin structure of the bicollinear magnetic order found in Fe$_{1+y}$Te \cite{li09a,bao09}.  (c) Stripe magnetic structure common to iron-pnictide parent compounds \cite{dela08,huan08a}.  (d) Corresponding locations of magnetic wave vectors for bicollinear (dark red) and stripe (magenta) orders in a single layer in reciprocal space (assuming twinned domains).  In (a)--(c), dashed lines indicate the unit cell.
}
\end{center}
\end{figure}

\begin{figure}[b]
\begin{center}
\includegraphics[width=0.99\columnwidth]{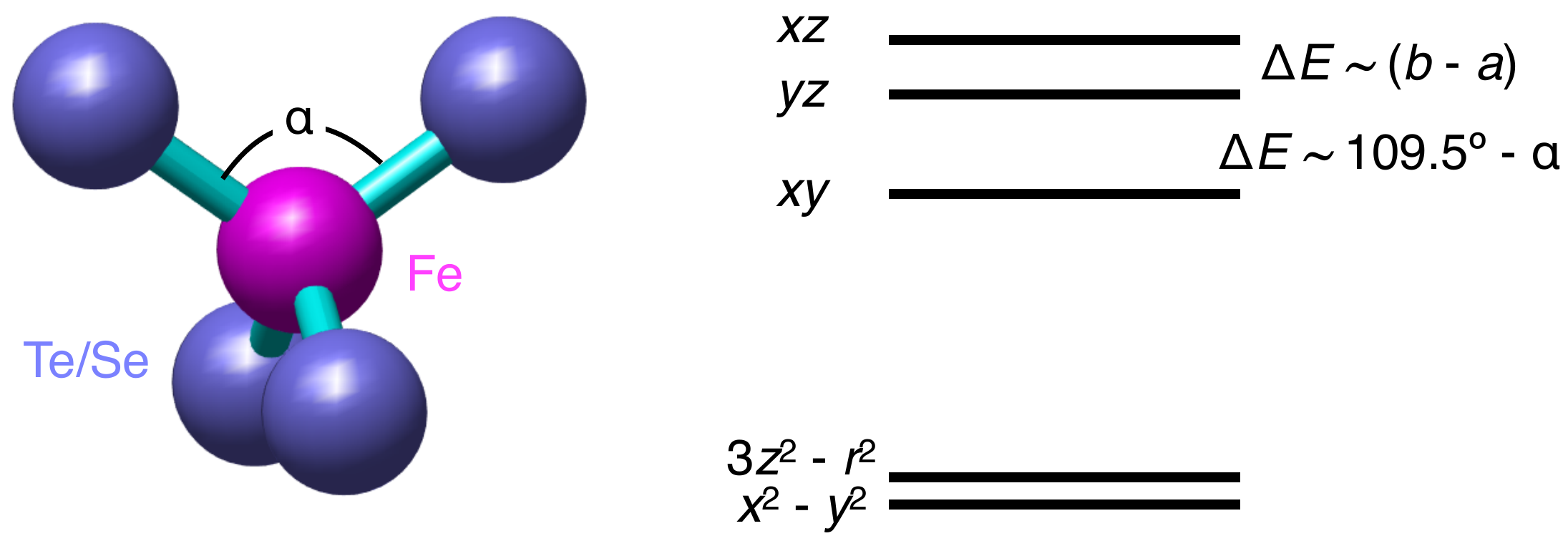}
\caption{\label{fg:tetra} 
Left:  Illustration of tetrahedral coordination of Fe by Te/Se.  Right: Crystal-field scheme of the relative energies of the Fe $3d$ states in the tetrahedral environment, including the effect of possible distortions, such as orthorhombic strain $(b-a)$ or distortion of the bond angle $\alpha$ from the ideal $109.5^\circ$.
}
\end{center}
\end{figure}

The structure of \fetese\ is fairly simple.  The Fe atoms form a square lattice [see Fig.~\ref{fg:struc}(a)] and are tetrahedrally coordinated by the Te/Se ligands located above and below the Fe plane (see Fig.~\ref{fg:tetra}), with identical layers simply stacked along the $c$ axis.  Because of the ligand arrangement, the unit cell contains 2 Fe sites, with $a\approx b \approx3.8$~\AA.  We will make use of this unit cell and the corresponding reciprocal space throughout this paper; however, when looking at the literature, it is important to be aware that many researchers assume a 1-Fe unit cell.

Antiferromagnetic order occurs in Fe$_{1+y}$Te \cite{li09a,bao09}, but with a pattern that is distinct from that commonly found in the iron pnictides \cite{dai15}.  These two spin structures, which are commonly labelled ``bicollinear'' and ``stripe'', are illustrated in Fig.~\ref{fg:struc}(b) and (c), respectively, and their characteristic wave vectors are plotted in Fig.~\ref{fg:struc}(d) (for the case of twinned magnetic domains).  As we will discuss, the characteristic wave vector evolves with Se substitution and with temperature, with low-frequency spin excitations characterized by the stripe wave vector developing in superconducting samples.

The coexistence of superconductivity and antiferromagnetic correlations raises interesting questions concerning the electronic structure.  Strong electronic correlation is universally accepted as a key feature in cuprate high-temperature superconductors, where it leads to a Mott insulator phase in the undoped parent materials. Hence, the importance of electronic correlation for the superconductivity and other properties of Fe-based superconductors (FeSC) have been at the focus of research ever since FeSC superconductivity was discovered \cite{Haule_NJP2009,YinHauleKotliar_NatMat2011,Yin_PRB2012,yin14,Georges_AnnuRev2013,SiYuAbrahams_NatRevMat2016}. The early electron-spectroscopic experimental investigations suggested FeSC might be in the weak to moderate correlation regime \cite{Yang_PRB2009,Yi_etal_Shen_PRB2009} and the early density functional theory (DFT) work also suggested weak correlation effects in these multi-band systems \cite{HanSavrasov_PRL2008,Subedi_PRB2008,Mazin_NatPhys2009,Mazin_Nature2010,ande11}. However, the DFT predictions were at odds with bad-metal behavior observed in the normal state of many FeSC materials and the evidence for strong correlations and massive spectral weight redistribution associated with magnetism in the Fe$_{1+y}$Te$_{1-x}$Se$_x$ family (11 family) \cite{Zhang_PRB2010}.

Subsequent theoretical work has established that the correlation strength in FeSC is substantial and that it is also material- and orbital-dependent. The strongest correlation effects were found in Fe$_{1+y}$Te, the chalcogenide family parent material, and for $d_{xy}$ and $d_{xz}/d_{yz}$ orbitals across all FeSC families \cite{Haule_NJP2009,YinHauleKotliar_NatMat2011}. The origin of strong correlations in these multiorbital metallic materials, which are not close to a Mott insulating state, was traced to the Hund's rule coupling (intra-atomic exchange) and these materials have been classified as Hund's metals \cite{Haule_NJP2009,YinHauleKotliar_NatMat2011,Yin_PRB2012,yin14,Georges_AnnuRev2013}. Hund's coupling is responsible for alignment of the spins of the $3d$ electrons on the same Fe site, which suppresses inter-orbital fluctuations; it can also lead to orbitally-selective strong correlations that depend on the average occupancy of the shell. The correlation effects generated by Hund's coupling explain bad-metal behavior of metallic systems with orbital multiplicity.

The $t_{2g}$ orbitals ($d_{xy}$, $d_{xz}$, and $d_{yz}$) sit at higher energy than the $e_g$ orbitals ($d_{x^2-y^2}$ and $d_{3z^2-r^2}$) due to the tetrahedral coordination of the Fe site. If each Fe atom sat within an ideal tetrahedron of ligands, the energies of the $t_{2g}$ states would be identical.  In reality, there are deviations from the ideal.  As noted in Fig.~\ref{fg:tetra}, an in-plane orthorhombic strain causes an energy splitting of the $d_{xz}$ and $d_{yz}$ orbitals, resulting in different degrees of hybridization with the ligands.  The resulting electronic anisotropy, observed in Fe$_{1+y}$Te and FeSe, is referred to as nematic order.  It is believed to be electronic in origin, although there are differing perspectives on whether magnetic or orbital-occupancy correlations are the driver \cite{fern14}.
Deviations of the bond angle from the ideal cause the energy of the $d_{xy}$ orbital to be lowered, and associated orbital-selective Mott behavior has been reported \cite{yi15,liu15}.   In \feytese, we observe temperature and composition dependent changes of lattice symmetry and bond angles, implying changes in the orbital occupancies that are also associated with variations in the spin correlations \cite{xu16}.  These both appear to be connected to nematic behavior, though a unique driver is not obvious.

The rest of the paper is organized as follows.  We discuss the phase diagram of \feytese\ in the next section.  In Sec.~3, we focus on the character of magnetic and charge correlations in Fe$_{1+y}$Te, which provided one of the earliest examples of orbital-selective correlation effects \cite{zali11}.  This will lead to a consideration of the short-range spin correlations observed by neutron scattering and their description by plaquette models, as discussed in Sec.~4.  In Sec.~5, we briefly mention temperature dependence of the spin fluctuations across the superconducting transition temperature, $T_c$, then discuss the variation of the spin correlations with temperature in the normal state in Sec.~6.  The effect of substitutions for Fe is covered in Sec.~7.  The impact of temperature-dependent changes in bond angles on orbital occupancy and magnetism are covered in Sec.~8.  A brief summary is presented in Sec.~9.

\section{Phase diagram}

\begin{figure}[b]
\begin{center}
\includegraphics[width=0.8\columnwidth]{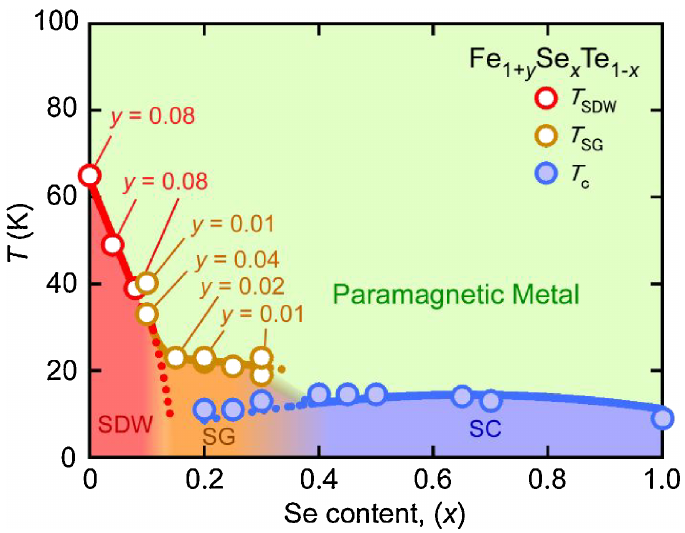}
\caption{\label{fg:phdiag} 
Phase diagram of \feytese\ with $y \ge0$ as a function of $x$ and $T$ from \cite{kata10}. The nominal Fe content, $y$, is zero unless specified otherwise. Blue circles: superconducting transition temperature, $T_c$; red circles: transition to long-range spin-density-wave order, $T_{\rm SDW}$; orange circles: onset of short-range spin-glass order, $T_{\rm SG}$. \copyright2010 The Physical Society of Japan.
}
\end{center}
\end{figure}

A phase diagram for \feytese\ is shown in Fig.~\ref{fg:phdiag} \cite{kata10}.  It has a superficial similarity to those of various iron arsenides \cite{hoso18}, with magnetic order at one end and a dome of superconductivity induced by chemical substitution \cite{yeh08,hsu08,fang08b,wen11}.  An important difference is that the doping in pnictides typically involves introduction of electrons (doping with Co for Fe) or holes (K for Ba), whereas Se and Te are isovalent.

A complication is that the Fe(Te/Se) layers are held together in the crystal only by weak Van der Waals forces. Crystallographic stability is improved if some amount of extra Fe atoms is incorporated between the layers \cite{liu11}, which is especially true in the case of Fe$_{1+y}$Te, where $0.02 < y < 0.18$ \cite{bao09,rodr11,stoc11,li09a,Martinelli2010,liu11,Koz_PRB2013}; in contrast, the excess Fe can approach zero in Fe$_{1+y}$Se \cite{mcqu09a}, and it is clear that excess Fe tends to suppress the superconductivity.   One really needs two composition axes to properly describe the phase diagram.

There are two features not properly represented in this phase diagram.  For one, there is evidence for bulk phase separation in the range $0.7\lesssim x\lesssim 0.1$ \cite{fang08b} which can be suppressed by epitaxial strain in thin films \cite{imai15}.  (Short-range segregation of Se and Te is observed for $x\sim0.5$ \cite{hu11,he11}, as the bond lengths for Fe-Se and Fe-Te are rather different \cite{louc10}.) For another, FeSe exhibits a structural transition at 90~K to an orthorhombic phase \cite{mcqu09b}, with a corresponding nematic electronic response and an absence of magnetic order  \cite{bohm17,cold18}.

Fe$_{1+y}$Te undergoes a structural transition from a tetragonal phase (space group $P4/nmm$) at high temperature to a phase in which the $a$-$b$ planes have an orthorhombic symmetry.  For $y\lesssim0.12$, the interlayer stacking corresponds to the monoclinic space group $P2_1/m$ \cite{li09a,Martinelli2010,zali12,McQueen_PRL2009}, while it changes to the orthorhombic space group $Pmmn$ for $y\gtrsim0.12$ \cite{rodr11,mizu12}.  The structural transition temperature drops from 75~K for $y\sim0.03$ to $\sim60$~K for $y\sim0.1$.  The antiferromagnetic order develops at or slightly below the structural transition.  Within the monoclinic phase, the order is bicollinear, as indicated in Fig.~\ref{fg:struc}(b), with modulation wave vector $(0.5,0,0.5)$ \cite{bao09,li09a,Parshall2012,zali12,rodr11}, but the order changes to helical incommensurate in the orthorhombic phase \cite{rodr11,stoc11,zali12}

A common feature of the structural transitions in chalcogenide, pnictide, and also cuprate superconductor families is an extreme sensitivity to disorder introduced by doping and off-stoichiometry. In the BaFe$_2$As$_2$ pnictide family, for example, transition metal substitution on the iron site at the 5--10\%\ level is sufficient to suppress the transition \cite{canf10,mart16}. In FeSe$_{1-x}$S$_x$, the orthorhombic phase is suppressed at $x \approx 0.15$ \cite{licc19}, and the situation is similar in the La$_{2-x}$Sr$_x$CuO$_4$ cuprate family \cite{axe94,kast98}. In \fetese, the structural transition to the monoclinic phase is suppressed by $x\approx0.1$. 

The unifying framework for understanding the doping-structure phase diagrams of layered crystalline materials with a symmetry-lowering transition, typically from a high-temperature $C_4$ tetragonal phase to a phase with $C_2$ symmetry in the $a$-$b$ plane, is provided by the anisotropic random field Ising model (ARFIM) \cite{ZacharZaliznyak2003}. The model was originally proposed to describe the sensitivity to disorder of the checkerboard charge order in half-doped layered perovskites, but it has broad applicability to electronic phases in cuprates, pnictides and chalcogenides \cite{zali12,Carlson_NatComm2011,Phillabaum_NatComm2012,Nie_PNAS2014}. At the effective theory level, the direction of atomic displacements selecting the symmetry breaking, {\it e.g.}\ lattice unit cell elongation along the $a$ (or $b$) axis, presents an Ising degree of freedom, while the strain mismatch between the two differently-distorted unit cells provides an effective ferromagnetic interaction. Doping-generated electrostatic disorder is the source of the random field, which precludes the symmetry breaking in the case of a purely two-dimensional system. For a material with weak inter-layer correlation of the distortions, the random potential strongly suppresses the symmetry breaking and results in a broad critical region in the phase diagram, with pre-existing domains of the low-symmetry phase and a percolation-type transition \cite{Carlson_NatComm2011,Phillabaum_NatComm2012,Nie_PNAS2014}. The structure-doping phase diagram of Fe$_{1+y}$Te$_{1-x}$Se$_x$ (Fig.~\ref{fg:phdiag}) fits naturally into this general ARFIM framework.

\section{From moment to moment in Fe$_{1+y}$Te}

The early band structure calculations predicted \fetese\ to be a metal with several bands crossing the Fermi energy \cite{HanSavrasov_PRL2008,Subedi_PRB2008,Mazin_Nature2010,Ma_PRL2009}. This qualitatively agreed with scanning tunneling spectroscopy \cite{Hanaguri_Science2010} and angle-resolved photoemission studies of FeTe \cite{Xia2009,ZhangFeng2010}, which found small electron and hole pockets near the corner, ${\bf k} = (0.5, 0.5)$, and the center, ${\bf k} = 0$, respectively, of the two-dimensional (2D) Brillouin zone (BZ) (recalling that we have chosen a unit cell containing 2 Fe atoms). While these findings revealed the existence of itinerant electrons, bulk resistivity measurements indicated either non-metallic or bad-metal behavior. At the same time, Curie-Weiss behavior of magnetic susceptibility of Fe$_{1+y}$Te suggested significant local magnetic moments, $\mu_{\rm eff} \approx 4 \mu_B$ (where $\mu_B$ is the Bohr magneton), and a rather small Curie-Weiss temperature,  $\Theta_{CW} \approx 190$ K \cite{Chen_PRB2009,Hu_etal_Petrovic_PRB2009,Liu_NatMat2010}. Thus, local moments and itinerant conduction electrons coexist in this material and the relation between them was uncovered by neutron scattering experiments, as we discuss next.

\subsection{Temperature dependence of the magnetic moment}

\begin{figure}[t]
\begin{center}
\includegraphics[width=0.6\columnwidth]{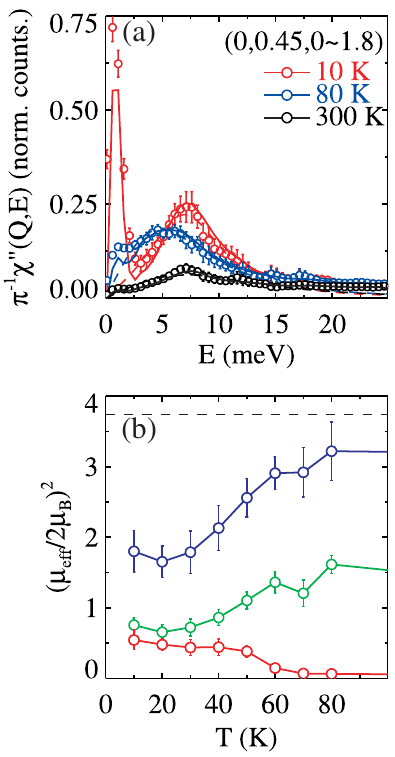}
\caption{(a) $\chi''(\bQ,E)$ as a function of energy in Fe$_{1+y}$Te ($y = 0.09$) for $\bQ = (0,0.45)$ at 10 K, 80 K and 300 K. Lines are fits to the imaginary dynamical susceptibility of a damped harmonic oscillator (DHO). (b) Square of the effective magnetic moment obtained by integrating the $S(E)$, as a function of temperature. Upper (blue) symbols show the total response, bottom (red) symbols are the Bragg contribution, green symbols are the quasi-elastic contribution. Reprinted with permission from Ref.~\cite{zali11}, \copyright2011 by the American Physical Society.
 }
\label{Fig:Tdep}
\end{center}
\end{figure}

The temperature dependence of magnetic scattering in Fe$_{1+y}$Te revealed that while the imaginary part of the local dynamical magnetic susceptibility does decrease with increasing temperature (Fig.~\ref{Fig:Tdep}a), a behavior that is expected in a local-moment system, the decrease is notably smaller than expected \cite{zali11}.
Indeed, for a local-moment system of spins $S$, the sum rule requires that the total spectral weight of the dynamical correlation function, ${\cal S}(Q, E)$, is conserved; that is, $\int {\cal S}(Q, E) d\bq dE = S(S+1)$ at all temperatures.  In Fe$_{1.1}$Te, the behavior is markedly different: it was observed that magnetic inelastic neutron scattering (INS) intensity significantly increases upon heating. The total magnetic spectral weight at 300~K yields $\mu_{\rm eff} \approx 3.6\,\mu_B$, close to the value of $\approx 3.9 \mu_B$ expected for $S = 3/2$, and in good agreement with the susceptibility data. At 10~K, however, the total spectral weight is roughly twice smaller, corresponding to $\mu_{\rm eff} \approx 2.7\,\mu_B$, and consistent with $S = 1$, as shown in Fig.~\ref{Fig:Tdep}(b). Thus, the overall picture is that of a temperature-induced change from local spins $S = 1$ at 10~K to $S = 3/2$ at 300~K. This can only occur as a result of an effective change by 1 in the number of localized electrons, with a corresponding change in the number of itinerant electrons, supporting a scenario of an orbital-selective localization (Mott transition).

The subsequent dynamical-mean-field theory (DMFT) calculations provided clear theoretical support for the orbital-selective electronic band decoherence in FeTe in this temperature range \cite{Yin_PRB2012}. Furthermore, a corresponding change in the electronic band structure near the Fermi energy was later observed by angle-resolved photoemission spectroscopy (ARPES) \cite{Liu_etal_Shen_PRL2013}. A decrease, amounting to the full disappearance of magnetic spectral weight, on cooling was also subsequently observed in the non-superconducting 122 parent material CaFe$_2$As$_2$, which is associated with the transition to a collapsed tetragonal phase \cite{Soh_etal_Goldman_PRL2013}. There, it can also be understood in terms of orbital-selective Mott physics (OSMP).

\subsection{Ferro-orbital bond-order wave in Fe$_{1+y}$Te}

The nature of the bicollinear antiferromagnetic phase observed in Fe$_{1+y}$Te for $y \lesssim 0.12$ \cite{li09a,Martinelli2010,liu11} presented an important problem for understanding the OSMP and distinguishing intertwined order parameters in the iron chalcogenide family. There are several aspects of this ordered phase that raised questions.  For one thing, the $(0.5,0)$ ordering wave vector of the antiferromagnetic structure is different from nesting wave vector $(0.5,0.5)$ predicted by calculations of the electronic band structure based on density-functional theory (DFT) \cite{HanSavrasov_PRL2008,Subedi_PRB2008}. The DFT nesting condition for FeTe is similar to the pnictide case, where the corresponding stripe order is, indeed, experimentally observed. This early failure of the itinerant electron description immediately brought strong correlation into relevance. An analysis of the corresponding local-spin model, however, indicated that quantum fluctuations actually select a different order, a double-{\bf Q} plaquette state, which preserves $C_4$ symmetry \cite{Ducatman_PRL2012,Ducatman_PRB2014}. This prediction appeared consistent with the short-range checkerboard structure of local dynamical correlations in the form of four-spin ferromagnetic (FM) plaquettes with antiferromagnetic (AFM) inter-plaquette correlations observed by neutron scattering in Fe$_{1.1}$Te \cite{zali11}. Such correlations, however, are incompatible with the bicollinear ground-state magnetic order, which has been firmly established by experiment \cite{li09a,Martinelli2010,liu11}.  Subsequent DFT analysis \cite{yin10} has shown that checkerboard, bicollinear and stripe antiferromagnetism are all in close competition for the ground state. This appeared consistent with the consensus that magnetoelastic coupling plays an important role in selecting the magnetic order in Fe$_{1+y}$Te.

\begin{figure}[t]
\begin{center}
\includegraphics[width=1.0\columnwidth]{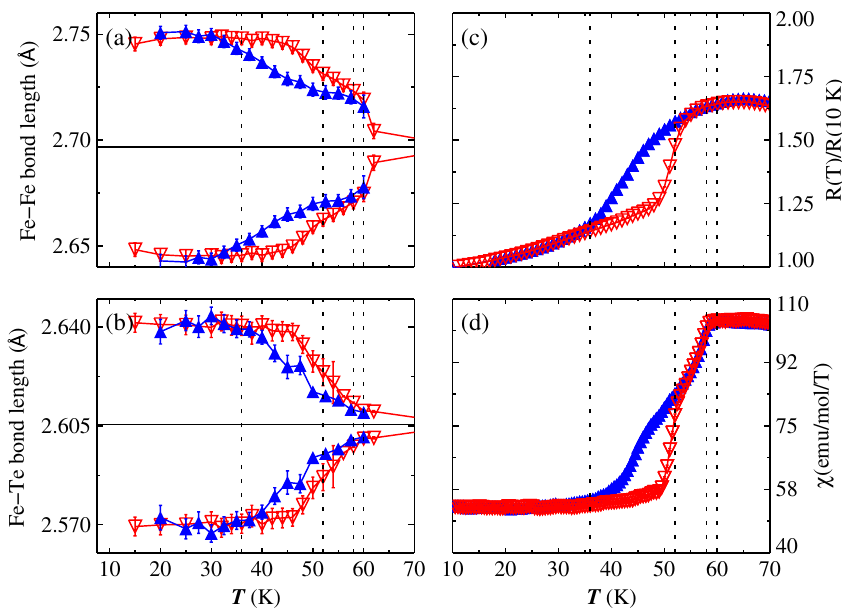}
\caption{Temperature dependence of Fe--Fe (a) and Fe--Te (b) bond lengths on cooling (blue closed) and warming (red open symbols). Resistivity (c) and magnetic susceptibility (d) measured on cooling (blue closed) and warming (red open symbols) in the same Fe$_{1.1}$Te.  The hysteresis in all quantities between 40 and 50~K indicates that the transition to the monoclinic phase, with modulated Fe--Fe bond lengths, is first order. Reprinted with permission from Ref.~\cite{Fobes_PRL2014}, \copyright2014 by the American Physical Society.
 }
\label{Fig:Tdep_BOW}
\end{center}
\end{figure}

While the importance of coupling to the lattice agrees with the magneto-structural nature of the first-order bicollinear magnetic ordering transition at $T_N \approx 70$--75~K in Fe$_{1+y}$Te for $y \lesssim 0.05$, the transition behavior becomes more complicated at larger $y$. With $y$ in the 5--10\%\ range, the transition splits into a sequence of second and first-order transitions involving different degrees of freedom \cite{zali12}. At $y \gtrsim 0.12$, well below any percolation phenomena, the bicollinear antiferromagnetism is replaced by a helical order \cite{bao09,rodr11,stoc11,li09a,Martinelli2010,liu11,Koz_PRB2013}. Even more important are the dramatic changes in bulk magnetic susceptibility and resistivity, which indicate emergence of metallic behavior in the ordered phase \cite{Chen_PRB2009}.  Figure~\ref{Fig:Tdep_BOW}(c) and (d) show examples of such changes for a crystal with $y=0.1$ \cite{Fobes_PRL2014} The single-crystal magnetic susceptibility decreases by nearly 30\%\ for all crystallographic directions, a behavior entirely unexpected for conventional antiferromagnetism, where transverse susceptibility is nearly temperature-independent below $T_N$.

These observations immediately point to the importance of both itinerant and strongly-correlated electronic characters in Fe$_{1+y}$Te.  The $\sim 30$\%\ decrease of magnetic susceptibility is consistent with the change of local magnetic moments from those corresponding to $S=3/2$ at high temperature to $S=1$ at low $T$ and, therefore, a delocalization of one of the three Fe $3d$ electrons, as inferred from the temperature dependence of magnetic dynamics [Fig.~\ref{Fig:Tdep}(b)] \cite{zali11}.  Orbital-selective electron delocalization also explains the emergence of metallic resistivity behavior and a Drude component in optical conductivity at low $T$ \cite{Chen_PRB2009,Hancock_PRB2010}. These behaviors are consistent with the DMFT results, which predict band coherence developing in FeTe at low temperature \cite{Yin_PRB2012}, and the ARPES observation of such coherence \cite{Liu_etal_Shen_PRL2013}.

\begin{figure}[t]
\begin{center}
\includegraphics[width=0.4\columnwidth]{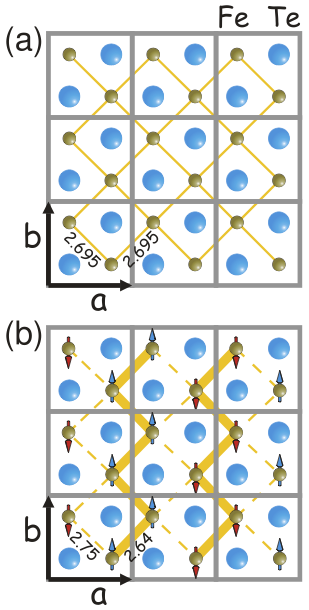}
\caption{(a) $P4/nmm$ unit cell of the square-lattice structure of a layer of FeTe projected onto the $a$-$b$ plane; (b)  bond-order wave (BOW) pattern of zigzag chains together with the bicollinear magnetic structure. Reprinted with permission from Ref.~\cite{Fobes_PRL2014}, \copyright2014 by the American Physical Society.
 }
\label{Fig:BOW}
\end{center}
\end{figure}

The puzzle of the bicollinear antiferromagnetic order in Fe$_{1+y}$Te was eventually solved by the combination of neutron diffraction and bulk resistivity and susceptibility measurements on single crystalline Fe$_{1.1}$Te samples \cite{Fobes_PRL2014}. For this stoichometry, as already mentioned, the first-order magnetostructral transition present at low $y$ is split into a sequence of transitions including the onset of incommensurate antiferromagnetism at $T_N \approx 58$~K and a broad hysteretic first-order transition in the 30--50~K range  \cite{zali12}, as illustrated by the resistance and susceptibility data in Fig.~\ref{Fig:Tdep_BOW}(c) and (d), respectively \cite{Fobes_PRL2014}. Neutron scattering measurements revealed the appearance of $(1,0,0)$ type Bragg reflections, which are forbidden in the high-symmetry tetragonal phase. The $(1,0,0)$ intensity quantifies the uniform displacement of one of the two Fe atoms in the unit cell, as shown in Fig.~\ref{Fig:Tdep_BOW}(a), which lowers the intra-unit-cell symmetry in the low-temperature phase. This displacement results in a substantial, $\approx 0.1$~\AA\ Fe-Fe bond disparity, which directly indicates the involvement of the orbital degree of freedom. As illustrated in Fig.~\ref{Fig:BOW}(b), the resultant formation of ferromagnetic zig-zag chains explains the bicollinear antiferromagnetic order.  This state is stabilized by the double-exchange mechanism induced by electron itinerancy, similar to that observed in the half-doped layered manganite, La$_{0.5}$Sr$_{1.5}$MnO$_4$ \cite{Sternlieb_PRL1996}.

The overall picture obtained from these experiments is that of an electronic instability in a correlated band of a Hund's metal, where the electronic coherence grows on cooling. This developing coherence and tendency to delocalize lead to ferro-orbital re-hybridization and formation of the bond-order-wave (BOW) pattern with zig-zag chains hosting the delocalized electrons. Consequently, the rising electron itinerancy leads to a decrease of the local magnetic moment and the emergent metallic behavior. The onset of the BOW also stabilizes the bicollinear magnetic structure, leading to the commensuration of the pre-existing magnetic order to the $(0.5,0)$ position \cite{zali12,Fobes_PRL2014}. The unidirectional nematic conductivity pattern implied by this scenario was subsequently confirmed by experimental measurements, where a marked resistivity anisotropy in the $ab-$plane was observed in the ordered phase of Fe$_{1+y}$Te \cite{Liu_etal_Uchida_PRB2015,Nakajima_PRB2015}.

\section{Local spin dynamics and plaquette correlations}

Despite the substantial ordered moment in the antiferromagnetic state of the parent chalcogenide, Fe$_{1+y}$Te, attempts to analyze the low-energy magnetic excitations in this system in terms of spin waves in a Heisenberg model have been unsatisfactory \cite{zali11,Lipscombe_PRL2011}. The inelastic magnetic scattering in this material is broad, diffuse, and typical of liquid-like short-range correlations, where the same well-defined pattern of local order persists in a broad range of time scales. This is not surprising in view of the presence of delocalized conduction electrons interacting with the system of atomic magnetic moments, which affect the nature of the spin excitations. The magnetic correlation patterns not only have to optimize the orbital overlap energy of localized valence electrons (spin superexchange), but also the hybridization energy with the wave functions of the delocalized (semi-)itinerant electrons \cite{Zaliznyak_PNAS2015}.

\begin{figure}[t]
\begin{center}
\includegraphics[width=0.95\columnwidth]{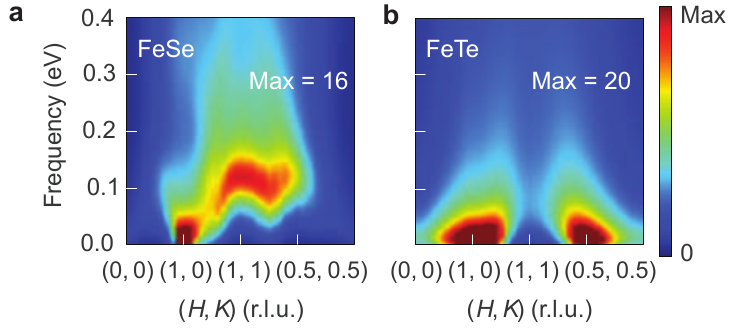}
\caption{\label{fg:dmft}  Magnetic dynamical structure factor for (a) FeSe and (b) FeTe calculated with a combination of DMFT and DFT approaches.  The maximum of the intensity color map (corresponding to dark red) is indicated at the upper right of each panel. Reprinted with permission from Ref.~\cite{yin14}, \copyright2014 Springer Nature.
}
\label{fg:dmft}
\end{center}
\end{figure}

A similar perspective is provided by {\it ab initio} calculations of the dynamical structure factor based on a combination of DMFT and DFT techniques \cite{yin14}, as shown in Fig.~\ref{fg:dmft}.  The calculations indicate a diffuse distribution of spectral weight, with weight shifted to higher energies in FeSe compared to FeTe.  The experimental measurements on FeSe are qualitatively consistent with this \cite{wang16b}.
 
\begin{figure}[t]
\begin{center}
\includegraphics[width=0.95\columnwidth]{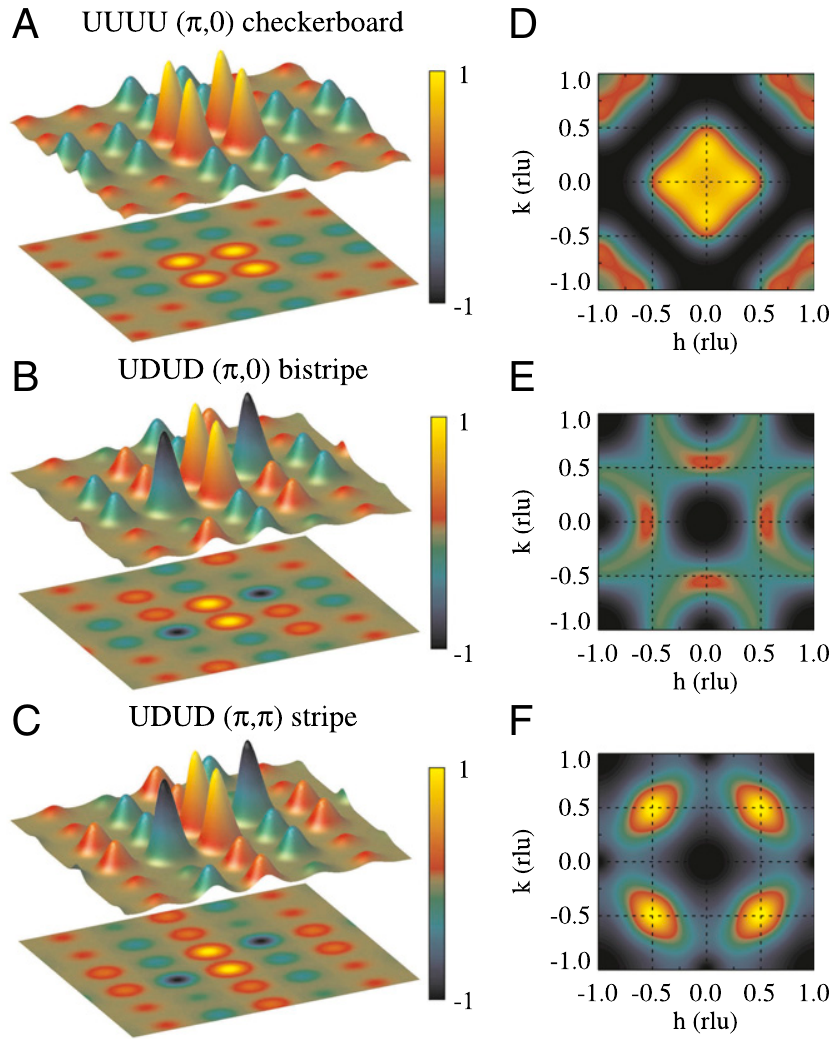}
\caption{Magnetization-density correlations in real and reciprocal space for plaquette spin-liquid models. (A-C) Height indicates absolute magnitude of the correlated magnetization density assuming isotropic atomic magnetization density. Color represents the relative amplitude of magnetic correlation, for interplaquette correlation length equal to $\sqrt{2}a$. (A) Ferromagnetic UUUU plaquette, with propagation wave vector $(0.5,0)$, corresponding to a checkerboard order.  (B) Antiferromagnetic UDUD plaquette, with wave vector $(0.5,0)$, corresponding to bicollinear order.  (C) Same plaquette as (B), but with ${\bf Q}=(0.5,0.5)$, corresponding to stripe order. (D-F) show the resulting scattering patterns in reciprocal space, averaged over all possible plaquette orientations to restore the macroscopic $C_4$ symmetry. Reprinted with permission from Ref.~\cite{Zaliznyak_PNAS2015}, \copyright2015 National Academy of Sciences.}  
\label{fg:plaquettes}
\end{center}
\end{figure}

Consistent with spin-liquid-like magnetic dynamics, it was found that the observed patterns of magnetic scattering in Fe$_{1+y}$Te can be very accurately described by a cluster model in which plaquettes of four up-up-up-up (UUUU) ferromagnetically co-aligned nearest-neighbor Fe spins emerge as a new collective degree of freedom, with short-range antiferromagnetic correlation between the neighboring plaquettes \cite{zali11}, as illustrated in Fig.~\ref{fg:plaquettes}A and D. The absence of magnetic scattering along the diamond described by ${\bf Q}=(\pm h,\pm k)=(\pm h,\pm(1-h))$ with $0\le h\le 1$ presents a clear fingerprint of the plaquette structure factor, ${\cal S}_p(\bQ) \sim |\cos(\pi (h+k)/2) \cos(\pi (h-k)/2)|^2$. Such ferromagnetic plaquettes are locally favored by Fe interstitials, a small density of which is present for $y > 0$ and which may act as condensation centers for the these correlations. This, however, appears insufficient to tilt the overall ground-state energy balance in their favor, as double exchange with itinerant electrons in the ferro-orbital BOW state favors the bicollinear long-range order. Thus, an unusual situation arises where the bicollinear static order coexists with the checkerboard-type dynamic correlations.
 
\begin{figure}[b]
\begin{center}
\includegraphics[width=0.85\columnwidth]{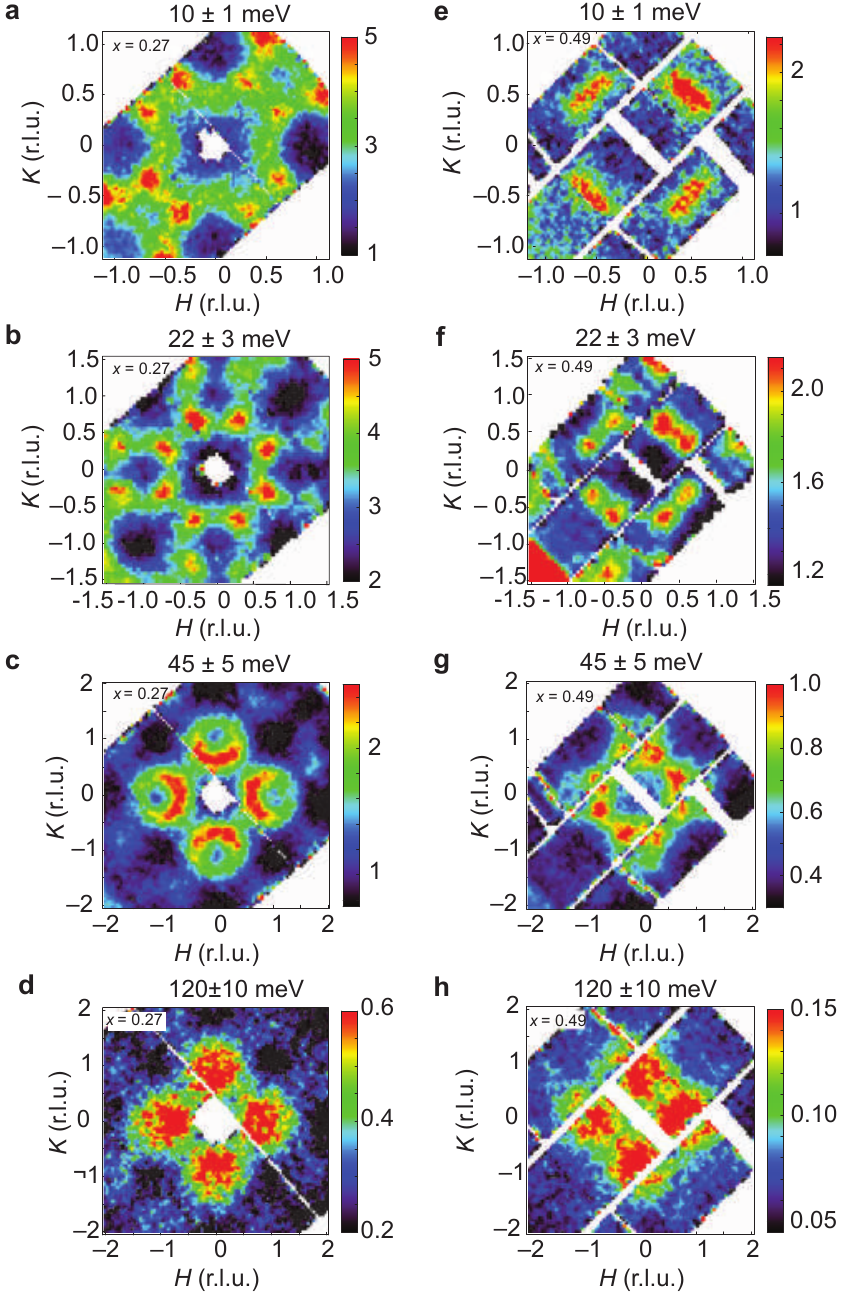}
\caption{Constant-energy slices of the inelastic magnetic scattering, projected onto the $(H,K)$ plane, for (a)-(d) nonsuperconducting Fe$_{1.04}$Te$_{0.73}$Se$_{0.27}$ at 5~K;  (e)-(h) superconducting FeTe$_{0.51}$Se$_{0.49}$ at 3.5 K.  The time-of-flight measurements were carried out with the sample $c-$axis parallel to the incident beam.  Excitation energy listed above each panel.  Reprinted with permission from Ref.~\cite{Lumsden_NatPhys2010}, \copyright2010 Springer Nature.
 }
\label{Fig:Lumsden_hkmaps}
\end{center}
\end{figure}

Whereas the nature of the local spin clusters that govern low-energy magnetic fluctuations in Fe$_{1+y}$Te is clearly not favorable for superconductivity, the situation changes with substitution of isoelectronic selenium (or sulfur) for tellurium in FeTe$_{1-x}$Se$_x$.  Neutron scattering shows marked changes of the low-energy inelastic magnetic scattering patterns with $x$, correlated with the appearance of superconductivity \cite{Zaliznyak_PNAS2015,Lumsden_NatPhys2010,Babkevich_PRB2011,Li_etal_Dai_PRL2010,xu12a}.  Figure~\ref{Fig:Lumsden_hkmaps} shows plots of inelastic magnetic scattering for two samples of \feytese, one with bulk superconductivity (righthand panels) and one without (lefthand panels).  At higher energies, the scattering patterns for the two samples look remarkably similar.  At the lowest energy, Fig.~\ref{Fig:Lumsden_hkmaps}(a) and (e), some differences appear.  The first thing to notice, however, is that in both cases, there is now strong scattering along ${\bf Q}=(h,1-h)$, making it clear that the spin correlations are quite different from the checkerboard model that works for Fe$_{1+y}$Te.

The significant width of the scattering patterns indicates that there is no need to give up on 4-spin plaquette models; rather, the solution is to change the choice of plaquette.  A suitable choice is the antiferromagnetic up-down-up-down (UDUD) plaquette indicated in Fig.~\ref{fg:plaquettes}B and C.  Applying a modulation wave vector of $(0.5,0)$ results in the bicollinear structure, as in Fig.~\ref{fg:plaquettes}B, while a modulation of $(0.5,0.5)$ leads to the stripe phase, as in Fig.~\ref{fg:plaquettes}C.  This is consistent with the suggestion that the corresponding magnetic states are close in energy \cite{yin10}.  It is apparent that these models are a reasonable starting point for describing the varying low-energy patterns demonstrated in Fig.~\ref{Fig:Lumsden_hkmaps}(a) and (e).

The initial application of these models was applied to neutron scattering measurements on a single-crystal sample of FeTe$_{0.87}$S$_{0.13}$, where two types of local order were found to co-exist and compete, changing their relative abundances as a function of temperature \cite{Zaliznyak_PNAS2015}. At high temperature, the scattering pattern had the checkerboard character, while the bicollinear correlations became important at low temperature.   These observations indicate an interesting example of liquid polymorphism, the coexistence of and competition between two distinct spin-liquid polymorphs and a liquid-liquid phase transformation between these states in a correlated electron system approaching superconductivity. 

It is important to note that the UDUD plaquette chosen in Fig.~\ref{fg:plaquettes} breaks $C_4$ rotational symmetry.  Of course, the local structural disorder from the Te and Se segregation effectively restores the symmetry on a large scale, and we must average over all possible plaquette orientations when modeling data.  Nevertheless, one can choose plaquettes with higher symmetry, as shown in \cite{xu16}, but the resulting scattering patterns give a much poorer description of the experimental data.  Hence, the observed magnetic scattering patterns provide evidence for broken symmetry, consistent with experimental observations that indicate a growing nematic susceptibility on cooling \cite{kuo16,xu16}.

\begin{figure}[b]
\begin{center}
  \includegraphics[width=0.99\columnwidth]{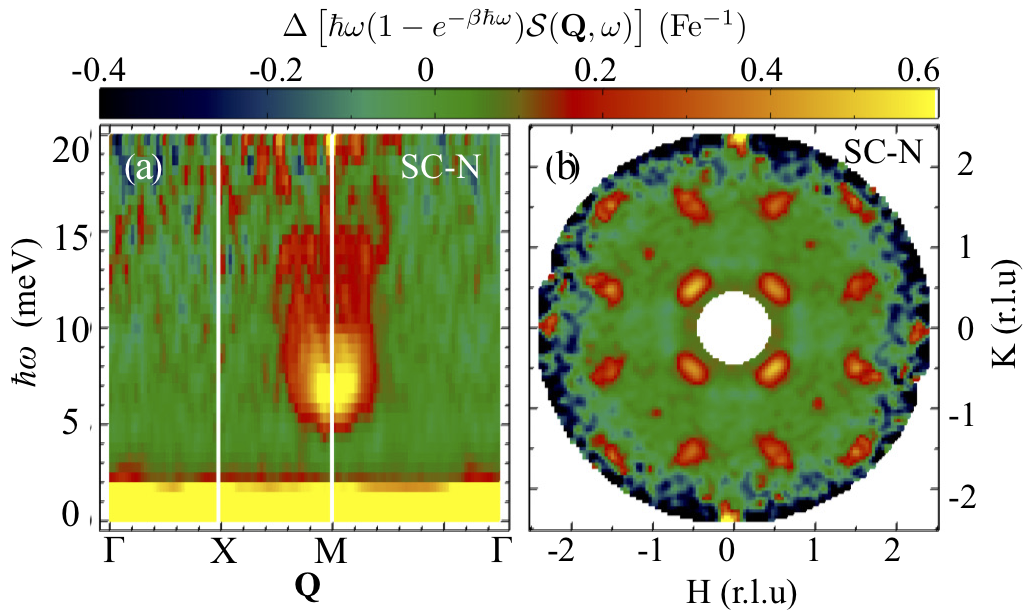}
\end{center}
\caption{Difference in INS for FeTe$_{0.6}$Se$_{0.4}$ ($T_c=14$~K) between superconducting  ($T=1.4$~K) and normal ($T=25$~K), plotted as $\chi''$ weighted by the energy transfer.  (a) energy dependence along high-symmetry directions; (b) {\bf Q} dependence for data integrated from 1 to 11 meV.  Reprinted with permission from Ref.~\cite{lein14}, \copyright2014 by the American Physical Society.} 
\label{fg:resonance}
\end{figure}

\section{Spin fluctuations and superconductivity}

In the superconducting state of optimally-doped \fetese, with $T_c\approx14$~K, magnetic scattering is suppressed at low frequency and an enhancement or resonance develops, centered on ${\bf Q}=(0.5,0.5)$ \cite{qiu09,lee10,xu10,argy10,babk10,mook10}.  As illustrated in Fig.~\ref{fg:resonance}, the resonance peak appears at $\hbar\omega\sim7$~meV for a sample of \fetese\ with $x=0.4$ and $T_c=14$~K \cite{lein14}.  As in the normal state, the resonance has an anisotropic shape in reciprocal space, being wider in the transverse direction.  The amplitude of the resonance is reduced by application of a magnetic field \cite{wen10} and an energy splitting, consistent with a singlet-triplet excitation, has been detected in a field of 14~T \cite{liu18}.

For comparison with the resonance energy, a superconducting gap $\Delta\approx2$~meV has been found in scanning tunneling microscopy \cite{hana10,sark17} and point-contact spectroscopy \cite{escu15} studies.  In contrast, ARPES measurements have indicated gaps of 1.7 and 2.5 meV on hole-like bands and 4.2 meV on the electron pocket \cite{miao12}, while gaps of 2.8 and 5.6 meV were found from optical conductivity work \cite{home15}.

In FeSe, the spin resonance also appears at the stripe wave vector, but the energy is scaled down to 4 meV, along with $T_c=8$~K \cite{wang16a}.  Towards the other end of the doping range of \feytese, where excess Fe tends to suppress superconductivity, recent work has shown that one can induce bulk superconductivity by annealing a sample in Te vapor.  While INS measurements on crystals with $x=0.1$ and 0.2 indicate a significant amount of bicollinear character in the normal-state magnetic scattering, the resonance signal is clearly centered at the stripe wave vector, but with considerable anisotropy in {\bf Q} width \cite{xu18}.

\section{$T$- and doping-dependence of spin correlations}

\begin{figure}[b]
\begin{center}
  \includegraphics[width=0.95\columnwidth]{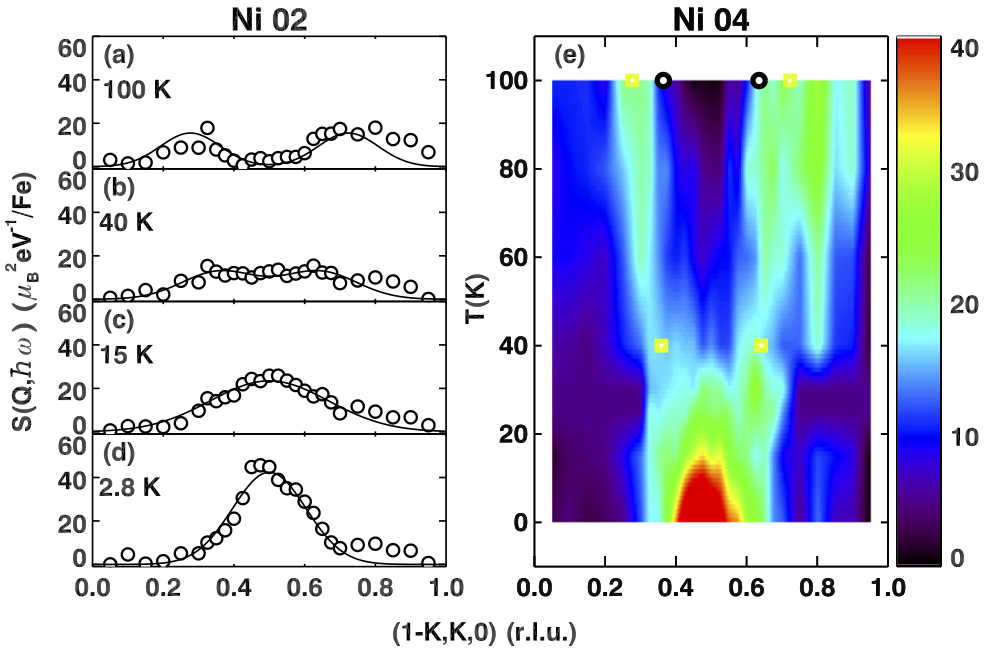}
\end{center}
\caption{Thermal evolution of the magnetic scattering at $\hbar\omega=5$~meV along ${\bf Q}=(1-K,K,0)$ for the Ni02 sample (see text) measured at (a) 100~K, (b) 40~K, (c) 15~K, (d) 2.8~K;  (e) related results for the Ni04 sample plotted as an intensity contour map in temperature--wave-vector space.  The data have been smoothed.  The yellow and black symbols in (e) denote the corresponding peak positions for the Ni02 sample (yellow squares) and for a superconducting Fe$_{1+\delta}$Te$_{0.35}$Se$_{0.65}$ sample~\cite{Zhijun_2011}.  Reprinted with permission from Ref.~\cite{xu12a}, \copyright2012 by the American Physical Society.
} \label{Xu2012:fig4}
\end{figure}

The redistribution of the magnetic spectral weight across $T_c$ is to be expected \cite{scal12a}; however, a change in the characteristic wave vector in the normal state was not.  An early observation of this effect is presented in Fig.~\ref{Xu2012:fig4}.  The samples are superconducting (SC) Fe$_{1-y}$Ni$_y$Te$_{0.5}$Se$_{0.5}$ with $y=0.02$ (Ni02, $T_c=12$~K) and $y=0.04$ (Ni04, $T_c=8$~K)  \cite{xu12a}.  Figure~\ref{Xu2012:fig4}(a) shows scans along ${\bf Q}=(1-k,k,0)$ for $\hbar\omega=5$~meV at several temperatures.  We previously noted the change in scattering along this direction with Se concentration in Fig.~\ref{Fig:Lumsden_hkmaps}(a) and (e).  Here we see a similar change, but due to varying temperature in the same sample.  Figure~\ref{Xu2012:fig4}(b) shows results on the Ni04 sample obtained at more temperature points.  At low temperature, the scattering is characterized by the stripe wave vector, but it crosses over to a different behavior above $\sim40$~K (well above $T_c$).

\begin{figure}[t]
\begin{center}
  \includegraphics[width=0.85\columnwidth]{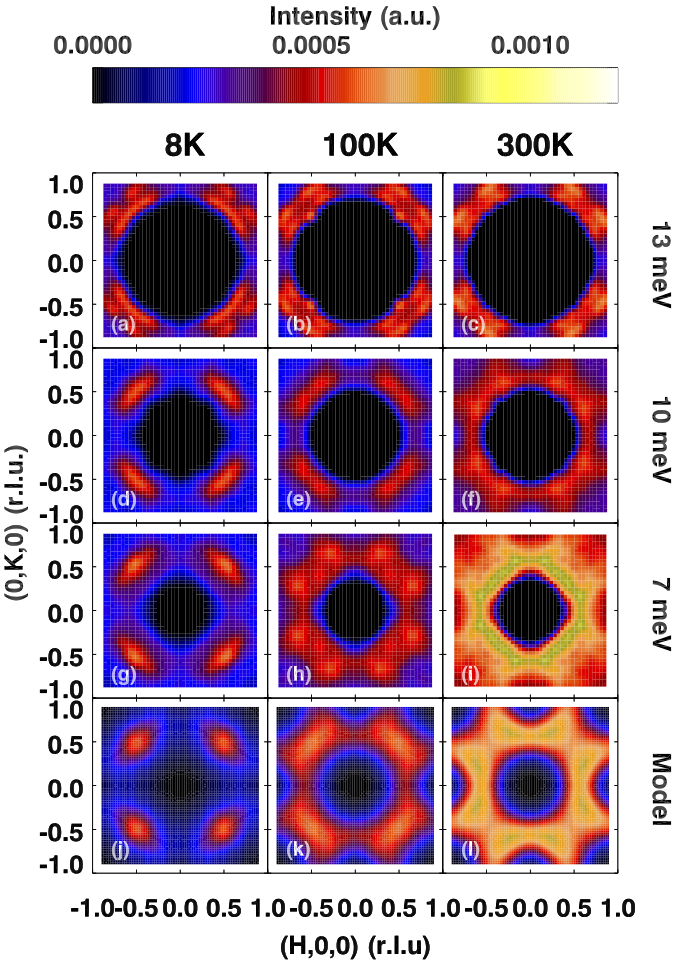}
\caption{Inelastic magnetic neutron scattering from the SC70 sample (see text) at energy transfers $\hbar\omega=$ 13~meV (a), (b), (c); 10~meV (d), (e), (f); and 7~meV (g), (h), (i).  The sample temperatures are 8~K (a), (d), (g); 100~K (b), (e), (h); and 300~K (c), (f), (i). All slices were taken with an energy width of 2~meV.  Measurements, covering approximately two quadrants, have been symmetrized to be 4-fold symmetric, consistent with sample symmetry.  Intensity scale is the same in all panels, but 13-meV data have been multiplied by 1.5 to improve visibility.  Black regions at the center of each panel are outside of the detector range. Panels (j), (k), (l) are model calculations simulating the 7-meV data, using the spin-plaquette model as described in the text, based on weakly correlated slanted UDUD spin plaquettes.  The inter-plaquette correlations used in the calculations correspond to (j) 100\% stripe, (k) 50\% stripe and 50\% bicollinear, and (l) 100\% bicollinear.  Reprinted with permission from Ref.~\cite{xu16}, \copyright2016 by the American Physical Society.}
\label{Zhijun2016:fig3}
\end{center}
\end{figure}

A 2D map of the low-energy magnetic scattering intensity from a SC sample, FeTe$_{0.3}$Se$_{0.7}$ (SC70), is shown in Fig.~\ref{Zhijun2016:fig3} \cite{xu16}.  At low temperature (8~K, left column) the magnetic excitations at 7 and 10~meV appear as ellipsoidal shapes near (0.5,0.5), elongated along the transverse directions. At 100~K, it is clear that the positions of the intensity maxima for 7~meV have shifted away from (0.5,0.5). With further heating to 300~K, the intensity forms a ``square'' ring structure that passes through the four equivalent (0.5,0)-type positions. The change of intensity distribution is much less pronounced at 10~meV and is hardly noticeable at 13~meV.  Measurements on a non-superconducting sample of Fe$_{1+y}$Te$_{0.55}$Se$_{0.45}$ reveal a similar pattern at 300~K, but with much less change on cooling to low temperature \cite{xu16}.

The bottom row of panels in Fig.~\ref{Zhijun2016:fig3} show calculations using the models of Fig.~\ref{fg:plaquettes}B and C, with Fig.~\ref{Zhijun2016:fig3}(j),(l), and (k) corresponding to stripe correlations, bicollinear correlations, and an average of the two.  Comparison with the 7-meV data suggests that the spin correlations change from stripe-like at low temperature to bicollinear at room temperature.  In contrast, the results on nonsuperconducting samples do not achieve the same stripe-like correlations at low temperature.  The experimental results are compatible with theoretical analyses in which low-energy spin excitations near (0.5,0.5) are directly related to the pairing mechanism in the  FeTe$_{1-x}$Se$_x$ compounds~\cite{Maier2009, Arita,Chubukov, TesanovicEPL, DHLee}.  At the same time, it is important to note that the correlations are always short-range.

\begin{figure}[t]
\begin{center}
\includegraphics[width=0.95\columnwidth]{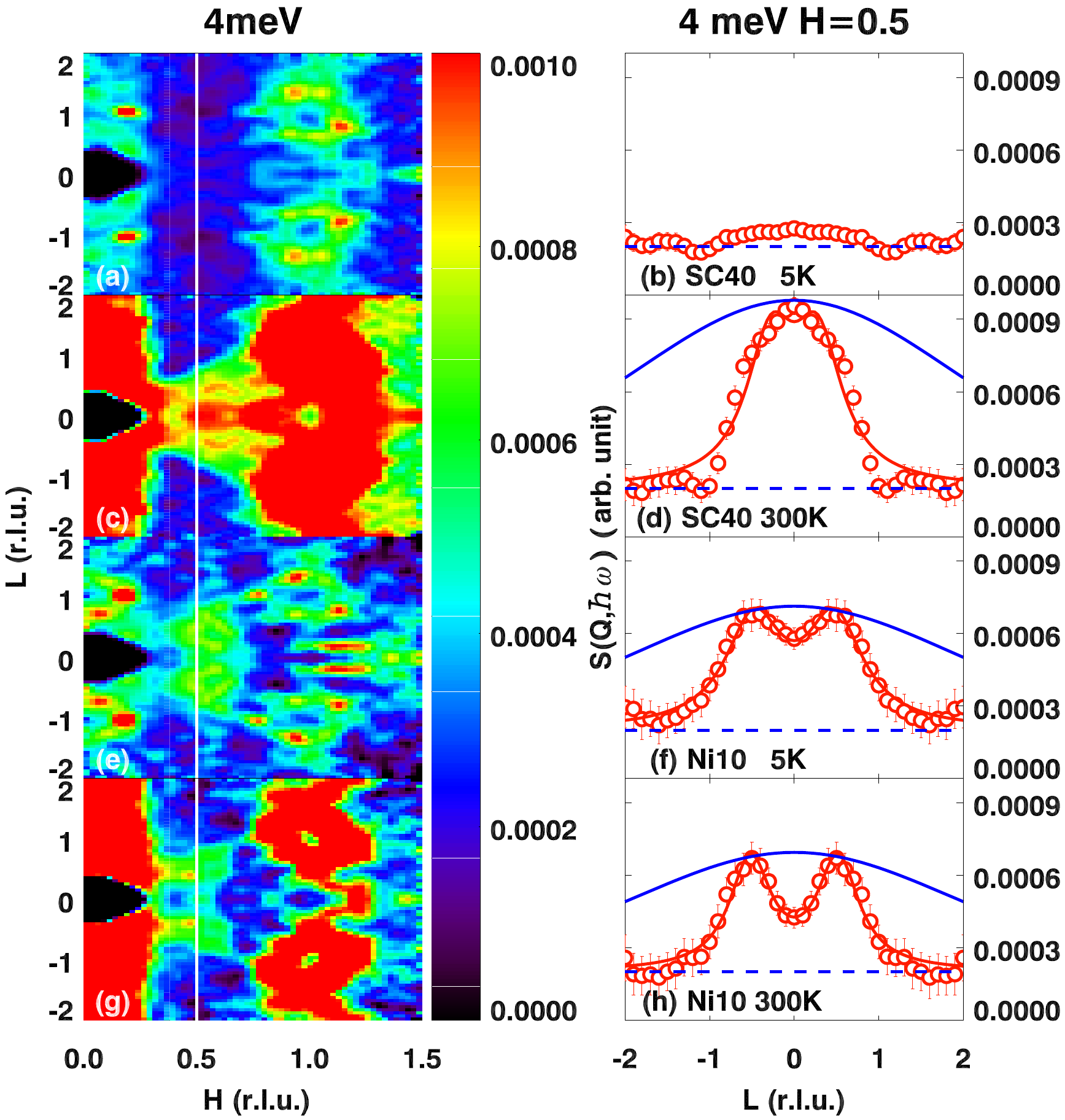}
\caption{Inelastic neutron scattering intensities in the $(H0L)$ plane measured at energy transfer $\hbar\omega= 4$~meV on the SC40 and Ni10 samples (see text). The intensities are scaled by the sample mass for better comparison. Left column are 2D intensity slices, and right column are linear intensity cuts along $(0.5,0,L)$. The q-width of the linear cuts is 0.05 r.l.u. along [100] direction. The panels are  (a) and (b): SC40 at 5~K; (c) and (d) SC40 at 300~K; (e) and (f) Ni10 at 5~K; and (g) and (h) Ni10 at 300~K.  The white line in the left panels at $H = 0.5$ shows where the $L$ cuts in the right column were taken. The dashed lines in right panels are estimated background values obtained from fitting around $(0.65,0,0)$.  The blue solid lines in (d), (f), and (h) are the magnetic form factor for an Fe$^{2+}$ ion scaled to the intensity maximum, and the red solid lines are fits to the data using two symmetric Lorentzians peaked at $L=\pm0.5$.  All slices were taken with an energy width of 2 meV. The error bars represent statistical error.  Reprinted with permission from Ref.~\cite{xu17}, \copyright2017 by the American Physical Society.}  \label{xu17:fig4}
\end{center}
\end{figure}

Another significant result concerns the dimensionality of the spin correlations.  For FeTe$_{1-x}$Se$_x$  compounds, the stripe-like correlations tend to be quite 2D \cite{qiu09}, whereas the bicollinear correlations have 3D character \cite{xu17}.  This leads to the surprising result that a superconducting sample starts off with 3D spin correlations at high temperature, and evolves to 2D correlations on cooling, before reaching $T_c$ \cite{xu17}.  To see this, the lefthand side of Fig.~\ref{xu17:fig4} shows measurements in the $(H0L)$ plane for $\hbar\omega=4$~meV.  The top two panels are for SC sample FeTe$_{0.6}$Se$_{0.4}$ (SC40). Panel (a) was measured in the SC state, where there is no spectral weight near $(0.5,0,L)$; warming to 300 K, panel (c), yields substantial intensity at $(0.5,0,0)$.  The variation of the intensity along $L$ is shown in (d); it varies much faster than does the Fe$^{2+}$ form factor, indicated by the blue line.  This indicates correlations along the $c$ axis between neighboring layers. 

Similar results are obtained from a nonsuperconducting sample of  Fe$_{0.9}$Ni$_{0.1}$Te$_{0.5}$Se$_{0.5}$ (Ni10), at both low (8~K) and high (300~K) temperatures. One can obtain a spin correlation length along the out-of-plane direction by fitting the data with two peaks at $L=\pm0.5$ (assuming antiferromagnetic interactions between Fe-planes), as indicated by the red line. Applying the same model to the SC40 sample, one obtains $\xi_c \sim 1.4(3)$~\AA, compared to the in-plane correlation length of $\xi_{ab}\sim1.7(4)$~\AA\ \cite{xu17}.

\section{Impact of substitution for Fe}

While superconductivity in  the ``11'' system can be tuned through the Se/Te ratio on the B site, substitution on the A site (Fe site) can also have interesting effects on the magnetic, and superconducting properties of the system. In this section we discuss the effects of A-site substitution on  FeTe$_{1-x}$Se$_x$.

In the case of BaFe$_2$As$_2$, substitution of Co or Ni for Fe gradually reduces the structural and magnetic transitions, and induces superconductivity beyond a threshold concentration;  Cu substitution also reduces the structural and magnetic transitions, but does not lead to superconductivity \cite{canf10}.  The case of Fe$_{1.1-z}$Cu$_z$Te is similar.  Doping Cu onto the Fe site results in a clear depression of the magnetic ordering temperature, and also a reduction of the spin gap, but no superconductivity \cite{wen12}. 

If we start with FeTe$_{1-x}$Se$_x$ ($x\sim 0.5$), we can follow the impact of A-site substitution on the superconducting state.  Having either excess Fe or transition-metal (Co, Ni, Cu) doping on the A site are found to depress superconductivity~\cite{Shipra2010,Nabeshima2011,wang15}. Excess Fe of about 5\%\ appears to be enough to drive the system completely nonsuperconducting \cite{Stock2012}. Among the transition metals, Cu seems to have the strongest effect and only $\sim 2\%$\ doping will completely suppress superconductivity. It takes between 4\%\ and 10\%\ of Ni doping to kill $T_c$, while in the case of Co doping, 10\%\ only reduces $T_c$ to $\sim 10$~K.

\begin{figure}[t]
\begin{center}
\includegraphics[width=0.9\columnwidth]{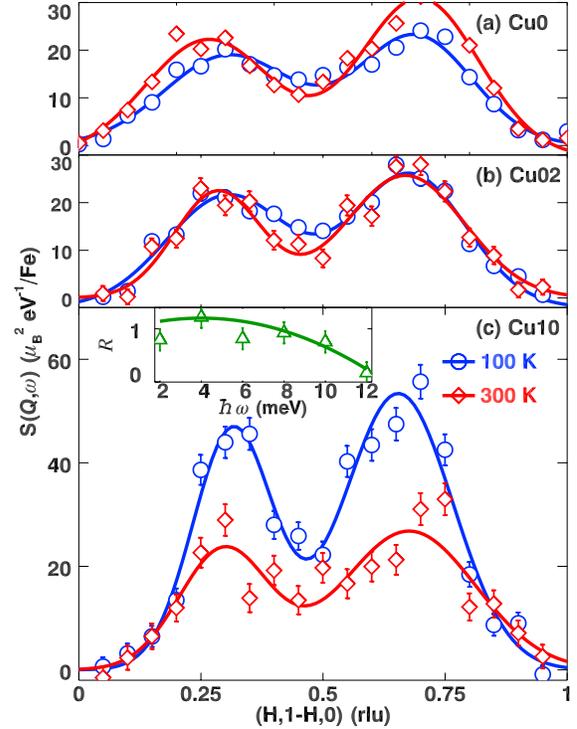}
\caption{Constant-energy scans of 6~meV through (0.5, 0.5) along [1$\bar{1}$0] direction at 100~K (blue circles) and 300~K (red diamonds) for (a) Cu0, (b) Cu02, and (c) Cu10 (see text).  Lines through data are fits with Gaussian functions. In the inset of (c) we plot the ratio (R) of enhancement on the 100-K integrated intensities (I$_{100K}$) to that of 300 K (I$_{300K}$) for these scans at different energies, with $R = (I_{\rm 100\,K} - I_{\rm 300\,K})/I_{\rm 300\,K}$. The line through the triangles is a guide to the eye.  Reprinted with permission from Ref.~\cite{wen13}, \copyright2013 by the American Physical Society.}  
\label{Wen2013:fig3}
\end{center}
\end{figure}

The magnetic properties can also be significantly affected.  When superconductivity is completely suppressed, the low energy spin excitations are also modified. Instead of suppressing the magnetic correlations as in the parent compound, A-site doping on SC compounds appears to eventually drive the system into the typical behavior of a nonsuperconducting ``11'' compound with (0.5,0) type spin correlations, as discussed in the previous section. However, there are subtle differences between the effects of excess Fe, Ni doping and Cu doping. With enough excess Fe, short-range magnetic order can be induced near (0.5,0,0.5), similar to that in the low-Se concentration nonsuperconducting (NSC) samples~\cite{xu10}. The low-energy excitations also show an enhancement upon heating (see Fig.~4 of Ref.~\cite{xu16}), likely a result of the conversion of ordered moment into low-energy spectral weight. On the other hand, no static magnetic order, neither long- nor short-ranged, has been found for the Ni and Cu doped samples. For a 10\%\-Ni-doped NSC sample, Ni10, the low-energy excitation intensity has very little temperature dependence and does not change much from 8~K to 300~K, as one can see in Fig.~\ref{xu17:fig4}. Cu doping seems to even enhance the low-energy spin excitations~\cite{wen13,wang15} when SC is completely suppressed.  Measurements on spin excitations at $\hbar\omega=6$~meV from three samples (no Cu doping: Cu0, 2\% Cu doping: Cu02, and 10\% Cu doping: Cu10) are shown in Fig.~\ref{Wen2013:fig3}. For both Cu0 and Cu02 samples, the intensities at $T=100$~K and 300~K area approximately the same; however, for the Cu10 sample, the excitations at $T=100$~K are almost twice as intense as those at 300~K.  The inset shows that this enhancement occurs for $\hbar \lesssim 10$~meV, with the difference between 100~K and 300~K disappearing above 12~meV.  Moreover, if one compares the normalized scattering intensities across different samples~\cite{wang15}, the low energy scattering intensity from the Cu10 sample is significantly higher than that from all other samples (undoped, Ni04, Ni10) at 100~K.

The peculiar behavior of the Cu10 sample may be explained if we propose that 10\%\ Cu-doping can lead to a (partial) localization of the itinerant electrons, and therefore enhances the  low-energy spectral weight. Consistent with electron localization, resistivity measurements show that the Cu10 sample has insulating behavior at low temperature, with resistivity three orders of magnitude higher than the undoped sample~\cite{wen13}. Similar effects have been reported in Cu-doped FeSe samples~\cite{Williams2009}. At high temperature (300~K), the localization effect diminishes and the scattering intensity of the Cu10 sample reverts back to that for the undoped or Ni doped samples.

Overall, we find that substitution of Fe with transition metals in superconducting  Fe$_{1+y}$Te$_{1-x}$Se$_x$  compounds suppresses superconductivity. At the same time, the magnetic correlations are not suppressed. Cu doping appears to be special, inducing a possible localization effect on the original itinerant electrons that results in an enhancement of low-energy spectral weight at low temperature.

\section{Thermal variation of bond angles}

\begin{figure}[t]
\begin{center}
\includegraphics[width=0.95\columnwidth]{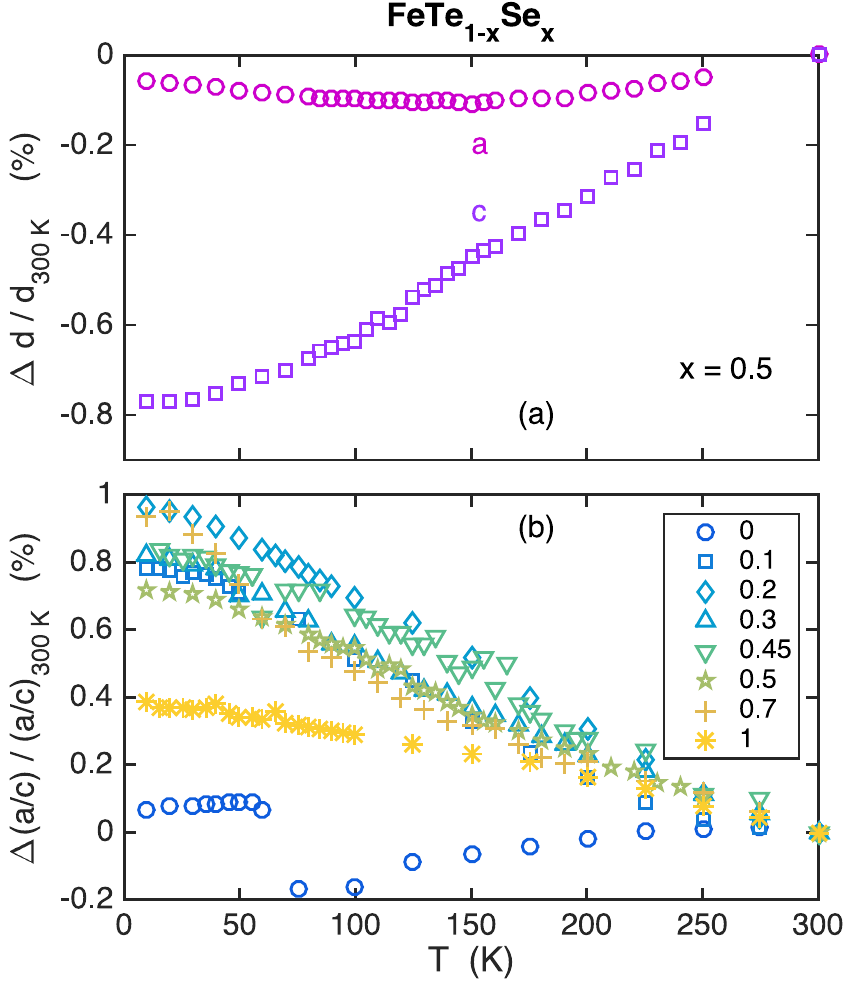}
\caption{\label{fg:latt} From \cite{xu16}.
(a) Change in $a$ and $c$ lattice parameters, normalized to 300~K, as a function of temperature for the $x = 0.50$ sample. Statistical uncertainties are comparable to the symbol size. (b) Change in $a/c$, normalized to 300~K, as a function of temperature for FeTe$_{1-x}$Se$_x$; the values of $x$ are noted in the symbol legend. The average of in-plane lattice parameters was used for a in the low-temperature phase of $x = 0$. Reprinted with permission from Ref.~\cite{xu16}, \copyright2016 by the American Physical Society.
}
\end{center}
\end{figure}

As noted in the introduction, the relative energies to the Fe $t_{2g}$ orbitals depend on the tetrahedral bond angle, as well as any in-plane distortion.  The effective magnetic couplings between neighboring Fe sites are sensitive to orbital energies and occupancies \cite{lee10,yin10,lee12}.  Could the change in magnetic correlations with temperature be connected with changes in bond angles?  

The Fe-Te and Fe-Se bonds tend to be relatively invariant \cite{louc10}, so changes in the structural $a/c$ ratio provides an indicator of changes in bond angle.  The temperature dependent changes in $a/c$ for a broad range of \fetese\ samples are presented in Fig.~\ref{fg:latt}.   The tetrahedral bond angle, indicated in Fig.~\ref{fg:tetra}, starts off much smaller than the ideal value; the temperature dependence is consistent with an increase in the bond angle, which brings the the $xy$ orbital energy closer to the $xz$ and $yz$.  This behavior is consistent with ARPES observations of an orbital-selective Mott transition on cooling \cite{yi15} and optical conductivity observations of enhanced in-plane Drude weight at low temperature \cite{home15}.


\section{Summary}

\begin{figure}[t]
\begin{center}
\includegraphics[width=0.9\columnwidth]{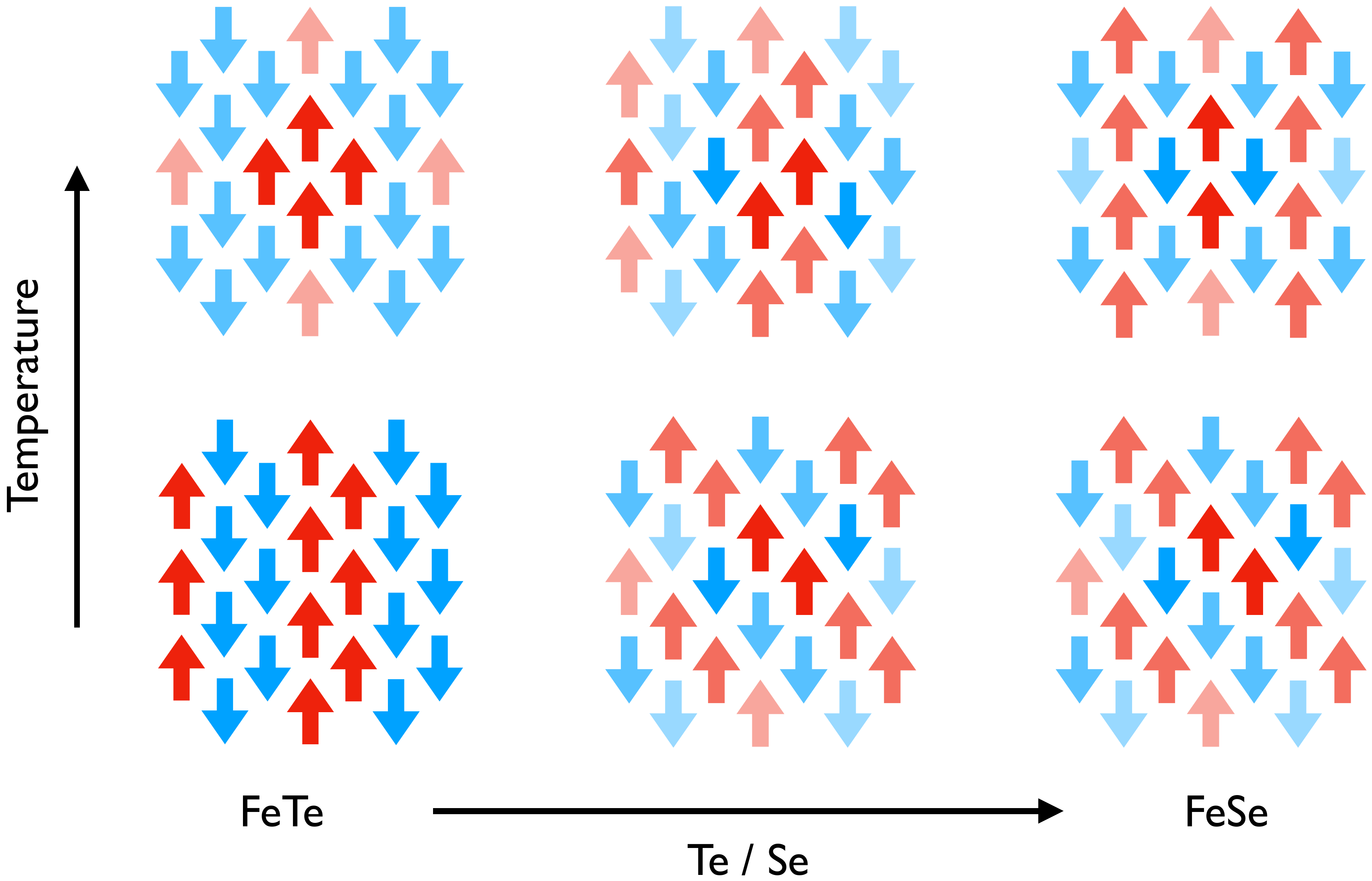}
\caption{\label{fg:sum} Schematic summary of the changes in magnetic correlations in \feytese\ with composition and temperature.
}
\end{center}
\end{figure}

Figure~\ref{fg:sum} presents a schematic summary of much of what neutron scattering has taught us about magnetic correlations in \feytese.  We have seen that there is a tendency for large magnetic moments but short-range magnetic correlations, with trade-offs between moment size and electronic conductivity.  Long-range magnetic order occurs in Fe$_{1+y}$Te, but with a characteristic wave vector inconsistent with superconductivity.  Selenium substitution tunes the magnetic correlations, but one still only achieves the dynamic stripe-like magnetic correlations, necessary for superconductivity, at fairly low temperature.  Differing orbital occupancies, including nematic effects, play important roles in determining the magnetic correlations.  A remaining challenge concerns the possible connection between the bulk magnetic correlations and the recently-reported topological surface states \cite{zhan19,rame19}.

\section{Acknowledgments}

JMT and IAZ are supported at Brookhaven by the Office of Basic Energy Sciences, U.S. Department of Energy, under Contract No.\ DE-SC0012704.  


\section*{References}
\bibliographystyle{iopart-num}
\bibliography{fe_sc,fetese,FeTereview,LNO,theory}

\end{document}